\documentclass[twocolumn,aps,showpacs,prb,tightenlines,amsmath,amssymb]{revtex4}
\usepackage{bm}
\usepackage{graphicx}              
\usepackage{amssymb}               
\usepackage{amsmath}                     
\usepackage{colordvi}
\usepackage{calrsfs} 
\newcommand{\bgreek}[1]{\mbox{\boldmath$#1$\unboldmath}}
\begin{document}   

\title{Hot-carrier transport and spin relaxation on the surface of topological insulator}
 
\author{P. Zhang}
\author{M. W. Wu}
\thanks{Author to whom correspondence should be addressed}
\email{mwwu@ustc.edu.cn.}
\affiliation{Hefei National Laboratory for Physical Sciences at
  Microscale and Department of Physics, 
University of Science and Technology of China, Hefei,
  Anhui, 230026, China} 
\date{\today}

\begin{abstract} 
We study the charge and spin transport under high electric field (up to several
kV/cm) on the surface of topological insulator Bi$_2$Se$_3$, where the
electron-surface optical phonon scattering dominates except at very low
temperature. Due to the spin mixing of conduction and valence bands, the 
electric field not only accelerates electrons in each band, but also leads to 
inter-band precession. In the presence of the electric field, electrons can transfer
from the valence band to the conduction one via the inter-band precession 
and inter-band electron-phonon scattering. 
The electron density in each band varies with the
electric field linearly when the electric field is strong. Due to the
spin-momentum locking, a transverse spin polarization, with the magnitude proportional to the momentum scattering
time, is induced by the electric field. The induced spin polarization depends on
the electric field linearly when the latter is small. Moreover, its magnitude is inversely
proportional to the temperature and is insensitive to the electron density at
high temperature. Our investigation also reveals that due to the large relative static
dielectric constant, the Coulomb scattering is too weak to establish a drifted
Fermi distribution with a unified hot-electron temperature in the steady state
under the electric field. After turning off the electric field in the steady state, the hot carriers cool down in a
time scale of energy relaxation which is very long (of the order of
100-1000~ps) while the spin polarization relaxes in a time scale of momentum scattering
which is quite short (of the order of 0.01-0.1~ps). 
 
\end{abstract}
%73.50.Fq  High field effect (conductivity) 
%75.70.Tj  Spin-orbit effects of thin films, surface and interface
%72.25.Rb  Spin relaxation and scattering
%72.25.-b  Spin polarized transport process
%71.10.-w  Theories of Many-electron systems

%85.75.-d  Spintronics
%75.76.+j  Spin transport 

\maketitle
\section{Introduction}
Topological insulators are a new class of materials that attract much interest
recently.\cite{fu106803,zhang438,hasan3045,qi1057,culcer860,xia398,beidenkopf,chen178,hsieh146401,heieh1101,kim459,hong2012,fiete845,analytis960,hasan55,tkachov} They are
band insulating in the bulk but conducting along the  
surface due to the gapless surface state. In three-dimensional strong
topological insulators, the surface state involves an odd number of massless Dirac cones
in the surface Brillouin zone and is protected by time-reversal
invariance.\cite{hasan3045,culcer860,qi195424,heieh1101,moore121306} For the simplest case with a single Dirac
cone, such as on the surface of Bi$_2$Se$_3$, Sb$_2$Te$_3$ and
Bi$_2$Te$_3$,\cite{zhang438,xia398,chen178,hsieh146401} the surface state near the Dirac cone can be described by the Rashba spin-orbit
coupling\cite{Rashba} and hence exhibits helicity.\cite{zhang438,liu045122,fu103} The metallic and helical surface state has
been proposed to have much potential for the application in spintronics.\cite{hasan3045,xia398} 

The spin helicity of surface state in topological
insulators has been experimentally measured by spin-angle resolved
photoemission spectroscopy.\cite{heieh1101,pan257004} However, it has
not been clearly evidenced by the electrical transport
experiment. This might be caused by the difficulty in seperating the
bulk and surface conduction as both of them are usually involved
simultaneously, and also the influence of the stray field from 
the ferromagnetic electrode when the spins are injected via the ferromagnetic
contact. To overcome these circumvents and confirm the spin-momentum
locking in the topological surface state, a scheme of transport experiment by
injecting spin polarized electrons into the topological surface state via
silicon has been proposed very recently.\cite{aristizabal} In spite of the
experimental difficulties, the understanding of charge and spin transport on the surface of topological
insulators is necessary. In fact, theoretically, this issue has been
preliminarily studied, with electric field being
small ($\sim$0.1~kV/cm,\cite{culcer860,culcer155457} under
 which the electrons are near equilibrium), the Fermi level 
located deep enough in the conduction 
band and the fully occupied valence band being
irrelevant.\cite{schwab67004,culcer155457,burkov066802,culcer860} 
Moreover, even in the microscopic study based on 
kinetic equations by Culcer {\sl et al.}, only the
electron-impurity scattering is considered.\cite{culcer155457} The
 electron-phonon scattering, which can be very
 important,\cite{zhu186102} and also the electron-electron Coulomb scattering
are not incorporated. It is revealed
that in the presence of driving by the electric field 
or diffusion by the density gradient, a transverse spin polarization is induced due to the spin-momentum
locking,\cite{schwab67004,culcer155457,culcer860} with 
the magnitude proportional to the electric field and momentum scattering
time.\cite{culcer155457,culcer860} Besides, due to the spin-momentum
locking again, the spin polarization relaxes in a time scale of momentum
scattering in the absence of an electric field.\cite{schwab67004,burkov066802} 

The transport on the surface of topological insulators
under large electric fields ($\sim$1~kV/cm), which can drive 
the carriers far away from the
equilibrium,\cite{weng245320,conwell} has been rarely investigated so
far. Here we perform this study on the surface of 
Bi$_2$Se$_3$ by means of the kinetic spin Bloch equation (KSBE)
approach,\cite{wu-review,cheng083704}
 with the electric field up 
to several kV/cm. Bi$_2$Se$_3$ is expected to have a 300 meV direct
band gap at the zone center\cite{zhang438,xia398} and is usually $n$-type due to the charged Se
vacancies. However, the electron density can be
adjusted by counterdoping with Ca\cite{wei201402,xia398,hor195208} or
partially substituting Bi with Sb to reduce Se
vacancies.\cite{analytis960,kulbachinskii15733} Particularly, employing Cd doping in
combination with a Se-rich growth condition, even a $p$-type Bi$_2$Se$_3$
can be obtained.\cite{ren075316} For the thin-layer
structure with a thickness $\sim$10~nm, the electron density is also
controllable by the gate voltage.\cite{kim459,hong2012} Due to the large relative static dielectric 
constant,\cite{kim459,culcer155457,richter619,butch241301} the
dominant scattering on the surface is the electron-surface optical
phonon scattering.\cite{zhu186102}

In this work, we take into account both the $n$-doped degenerate and
intrinsic nondegenerate cases before turning on the electric field, by including both
the conduction and valence bands. Our study reveals that due to the joint effects of the driving of the
electric field and the inter-band precession as well as inter-band electron-phonon scattering,
electrons can be transferred from the valence band to the conduction
one. This effect is more pronounced when the scattering is weak and  
the electric field is strong. Moreover, the variation in electron density for each band is linear in the
electric field when the latter is strong for both the $n$-doped and
intrinsic cases. The induced spin polarization is linear in the electric field when
the latter is small and deviates from the linear relation when the latter
becomes large enough. Besides, at high temperature, the induced spin
polarization is sensitive to the temperature but insensitive to the electron density. We also
find that the electron-electron Coulomb scattering is too 
weak to establish a drifted Fermi distribution with a unified hot-electron temperature under the electric field. After turning off the electric field, the hot carriers cool down in a
time scale of 100-1000~ps while the spin polarization relaxes in a time scale of
momentum scattering, which is of the order of 0.01-0.1~ps.

This paper is organized as follows. In Sec.~II, we introduce the model and KSBEs. In
Sec.~III, we analytically solve the KSBEs under a small electric field in the presence
of electron-impurity and electron-phonon scatterings, with the latter treated in
the elastic scattering approximation. In Sec.~IV we present the numerical results from full
calculation based on the KSBEs. We summarize in Sec.~V.

\section{Hamiltonian and KSBEs}
We set the $z$-axis along the $\langle 001\rangle$ direction of Bi$_2$Se$_3$. In the
collinear spin space spanned by the eigenstates of $\sigma_z$, $\{|\uparrow\rangle,|\downarrow\rangle\}$, the
single-electron Hamiltonian describing the low-energy $(001)$ surface state
around the $\Gamma$ point has the Rashba spin-orbit coupling\cite{Rashba} form\cite{zhang438,liu045122,fu103} 
\begin{align}
H_0=\hbar v_f({\bf k}\times \hat{\bf z})\cdot{\bgreek \sigma},
\label{h0}
\end{align}
where $v_f\approx 5\times 10^5$~m/s is the Fermi velocity [$v_f$ is experimentally
measured to vary from $2\times 10^5$~m/s (Ref.~\onlinecite{beidenkopf}) to
$6\times 10^5$~m/s (Ref.~\onlinecite{zhang438})], ${\bf k}$ is the
two-dimensional electron momentum and ${\bgreek \sigma}$ is the vector
composed of the Pauli matrices for spin. The eigenstates of $H_0$, $|{\bf k}\pm\rangle=\frac{1}{\sqrt{2}}(\pm
ie^{-i\theta_{\bf k}}|\uparrow\rangle+|\downarrow\rangle)$ with
$\theta_{\bf k}$ representing the polar angle of ${\bf
  k}$, correspond to the conduction
and valence bands with linear dispersion $\varepsilon_{{\bf k}\pm}=\pm\hbar
v_fk$. The interaction Hamiltonian $H_I$ includes the electron-impurity,\cite{culcer155457}
electron-surface optical phonon\cite{zhu186102} and electron-electron Coulomb
interactions. 

The KSBEs\cite{wu-review,cheng083704} are applied to study the charge and spin dynamics on the
surface of Bi$_2$Se$_3$ in the weak scattering regime where $v_f  \langle k\tau_k\rangle \gg 1$ ($\tau_k$ is the
momentum scattering time). In the helix spin space constructed by $\{|{\bf
  k}+\rangle,|{\bf  k}-\rangle\}$,\cite{cheng083704} the KSBEs read\cite{wu-review,cheng083704}   
\begin{align}\nonumber 
  &\partial_t\rho_{\bf k}(t)+i[v_fk\sigma_z,\rho_{\bf
    k}(t)]-(eE/\hbar)\partial_{k_x}\rho_{\bf k}(t)\\ & -(eE/\hbar)[U_{\bf
    k}^\dagger\partial_{k_x}U_{\bf k},\rho_{\bf k}(t)]+\partial_t\rho_{\bf
    k}(t)\big|_{\rm{scat}}=0. 
\label{ksbe}
\end{align}
Here $\rho_{\bf k}(t)$ is the $2\times 2$ density matrix for electrons with
momentum ${\bf k}$ in the helix spin space. The diagonal elements of $\rho_{\bf k}(t)$, $\rho_{{\bf
    k}++/--}(t)\equiv f_{{\bf k}+/-}(t)$, stand for the distributions of 
electrons in the conduction and valence bands respectively, and the off-diagonal
ones, $\rho_{{\bf k}+-}(t)=\rho_{{\bf k}-+}^\ast(t)$, characterize the inter-band
coherence. The second term on the left-hand side of Eq.~(\ref{ksbe}) is the
coherent term. The third and fourth terms are contributed by the
electric field along the $x$-axis, ${\bf E}=E\hat{\bf x}$. The third term
accelerates electrons in each band, while the fourth term, with  
\begin{align}
  U_{\bf k}=\frac{1}{\sqrt{2}}\left(
    \begin{array}{cc}
      ie^{-i\theta_{\bf k}} & -ie^{-i\theta_{\bf k}}\\
      1 & 1
    \end{array}
  \right)
  \label{uk} 
\end{align}
and hence $U_{\bf k}^\dagger\partial_{k_x}U_{\bf k}=\frac{i\sin\theta_{\bf
    k}}{2k}(1-\sigma_x)$, leads to the inter-band precession. This
electric-field--induced inter-band precession  originates from 
 the spin mixing in the conduction and valence bands. In fact, 
this effect also exists
in graphene where the pseudo-spins are mixed in the conduction and valence
 bands, as revealed by Balev {\sl et al.}.\cite{balev165432} However, in their
study with a low electric field $\sim$0.01~kV/cm,  the inter-band
coherence and hence naturally the precession term are neglected in their
kinetic equations.\cite{balev165432} Here we retain this term in the KSBEs
and will show it influences the transport properties markedly.
The last term on the left-hand side of Eq.~(\ref{ksbe}) is the scattering term, explicitly
given in Appendix~\ref{ap1}. Due to the large relative static dielectric 
constant $\kappa_0$ of Bi$_2$Se$_3$ (we typically set $\kappa_0=100$
following Refs.~\onlinecite{kim459} and \onlinecite{culcer155457}
while $\kappa_0\approx 50\sim 200$ as given in the literature\cite{richter619,butch241301}) and
the low energy of surface optical phonons
($\hbar\omega_0=7.4$~meV), the electron-electron and electron-impurity 
scatterings are less important than the electron-phonon
scattering in a large temperature region.\cite{giraud245322,hatch241303,butch241301} 

By solving the KSBEs in the presence of
electric field, we can investigate the 
charge and spin transport properties on the surface of topological
insulator. The electron and hole densities in the two
bands are $n_e=\sum_{\bf k}f_{{\bf k}+}$ and $n_h=\sum_{\bf k}(1-f_{{\bf k}-})$,
respectively. The spin polarization in the collinear spin space, induced by  
the electric field, reads
\begin{align}
{\tilde {\bf S}}=\sum_{\bf k}{\tilde{\bf S}}_{\bf
  k}=\frac{\hbar}{2}\sum_{\bf k}{\rm Tr}[{\tilde \rho}_{\bf k}{\bgreek
  \sigma}]
\label{sd}
\end{align}
where ${\tilde \rho}_{\bf k}=U_{\bf k}\rho_{\bf k}U_{\bf k}^\dagger$ is the electron density matrix in the collinear spin
space. Due to the Rashba spin-orbit coupling [Eq.~(\ref{h0})], the
charge current density along the $x$-axis under the electric field reads
\begin{align}
{\tilde j}_x=ev_f\sum_{\bf k}{\rm  Tr}[{\tilde \rho}_{\bf
  k}{\sigma}_y]=2ev_f\tilde{S}_y/\hbar,
\label{current}
\end{align}
exactly proportional to the induced spin polarization. Starting
from the steady state established by the electric field, we can study
the cooling of carriers and the 
relaxation of spin polarization by solving the KSBEs in the absence of the electric
field. In the following, we carry out these studies first analytically under a
low electric field and then numerically in the large electric field regime.

\section{Analytical study with elastic scattering at low electric
  field}
\label{ana}
Under a small electric field ($eE\langle\tau_k\rangle\ll\hbar \langle
k\rangle$), the KSBEs can be solved analytically with only the elastic
scattering, e.g., the electron-impurity scattering. Due to the low
energy of surface optical phonons, we can also
incorporate the electron-phonon scattering by elastic scattering 
approximation. Therefore, with both the electron-impurity and
electron-phonon scatterings, the scattering term in Eq.~(\ref{ksbe}) 
can be simplified from Eq.~(\ref{scat-app}) as
\begin{align}\nonumber
 &\partial_t\rho_{\bf k}(t)\big|_{\rm scat}=\frac{k}{8\pi\hbar^2 v_f}\int_0^{2\pi}d\theta_{{\bf
      k}^\prime}\big[n_i|D_{{\bf k}-{\bf
        k}^\prime}|^2+(2N_0+1)\\\nonumber &
\hspace{1 cm}\times|g_{{\bf k}-{\bf k}^\prime}|^2\big]\Big\{2\big[1+\cos(\theta_{\bf k}-\theta_{{\bf
        k}^\prime})\big]\big[\rho_{\bf k}(t)-\rho_{{\bf
        k}^\prime}(t)\big]\\ &\hspace{1 cm}+i\sin(\theta_{\bf k}-\theta_{{\bf
      k}^\prime})\big[\rho_{\bf k}(t)-\rho_{{\bf
      k}^\prime}(t),\sigma_x\big]\Big\}\Big|_{k^\prime=k}. 
  \label{scat_ei}
\end{align}
Here $|D_{{\bf k}-{\bf k}^\prime}|^2$ and $|g_{{\bf k}-{\bf k}^\prime}|^2$ are the
electron-impurity (refer to Appendix~\ref{ap1}) and electron-phonon scattering matrix
elements respectively, $n_i$ is the impurity density, and $N_0$ is 
the number of optical surface phonons. Explicitly, 
\begin{align}
|g_{\bf   q}|^2\approx (\lambda_1+\lambda_2q)^2\hbar/(2M{\frak A}\omega_0),
\label{ep-me}
\end{align}
where $M$ is the ion mass, ${\frak A}$ is the primitive cell area [$\hbar^2/(M\frak A)=4\times
10^{-3}$~meV], $\lambda_1\approx 5$~eV$\cdot$nm and
$\lambda_2\approx 1.6$~eV$\cdot$nm$^2$.\cite{zhu186102} In our study, the
largest $q$ involved is of the order of 0.1-1~nm$^{-1}$ and therefore
the momentum dependence of $|g_{\bf q}|^2$ is weak in the
presence of the large constant $\lambda_1$. It is noted that in
Eq.~(\ref{scat_ei}) the contribution with $|\theta_{\bf k}-\theta_{{\bf 
    k}^\prime}|=\pi$ is zero, indicating the absence of backscattering in the
same band. 
 
By performing the Fourier transformation on Eq.~(\ref{ksbe}) with respect to the polar angle $\theta_{\bf k}$ and defining ${\bf S}_k^l(t)\equiv\frac{1}{2\pi}\int_0^{2\pi}d\theta_{\bf
  k}e^{-il\theta_{\bf  k}}{\rm Tr}[\frac{\hbar}{2}\rho_{\bf k}(t){\bgreek \sigma}]$, one has
\begin{align}\nonumber
&\partial_t{\bf S}_k^l(t)+2v_fk{\bf
   S}_k^l(t)\times{\hat{\bf z}} +{\bf
      S}_k^l(t)/\tau_{1k}^l-i{\bf S}_k^l(t)\times\hat{\bf x}/\tau_{2k}^l\\ &
-eE\delta_{l,\pm1}\Delta_k^1\hat{\bf z}/4  -ieEl\delta_{l,\pm1}\Delta_k^0k^{-1}\hat{\bf y}/4=0. 
\label{ksbe-ana}
\end{align}  
Here $\Delta_k^0=f_{{\bf
    k}+}(0)-f_{{\bf k}-}(0)$, $\Delta_k^1=\partial_k\Delta_k^0$ and the
  momentum scattering rates
\begin{align}
  \frac{1}{\tau_{1k}^l}=&\frac{k}{4\pi\hbar^2
   v_f}\int_{0}^{2\pi}d\theta[n_i|D_{\bf q}|^2+(2N_0+1)|g_{\bf q}|^2] \nonumber\\
 & \times (1+\cos\theta)(1-\cos l\theta),\label{st1}\\\nonumber 
 \frac{1}{\tau_{2k}^l}=&\frac{k}{4\pi\hbar^2
   v_f}\int_{0}^{2\pi}d\theta[n_i|D_{\bf q}|^2+(2N_0+1)|g_{\bf q}|^2]
 \\ & \times\sin\theta\sin l\theta.
\label{st2}
\end{align}
To obtain Eq.~(\ref{ksbe-ana}), we have replaced $\rho_{\bf k}(t)$ in
the terms proportional to $E$ in Eq.~(\ref{ksbe}) by the initial
equilibrium state $\rho_{\bf k}(0)=\mbox{diag}\{f_{{\bf k}+}(0),f_{{\bf
    k}-}(0)\}$ when $E$ is small, and also assumed $|D_{\bf q}|^2$ to depend only
on the magnitude but not the direction of ${\bf q}$
($q=2k\sin\frac{\theta}{2}$) by treating the screening in the
  long-wavelength and static limit for analytical feasibility (refer to Appendix~\ref{ap1}). Here $f_{{\bf
    k}\pm}(0)=1/\{\exp[\beta(\varepsilon_{{\bf k}\pm}-\mu)]+1\}$ is the Fermi
distribution, where $\beta=1/(k_BT)$ and $\mu$ is the chemical potential. In
both the degenerate and nondegenerate limits, $|D_{\bf q}|^2\propto k^{-2}$ approximately 
(refer to Appendix~\ref{ap1}) and hence the momentum scattering rate limited by the
electron-impurity scattering is $\propto k^{-1}$. However, for the
electron-phonon scattering, with $|g_{\bf q}|^2\approx |g_{0}|^2\equiv |g_{{\bf
    q}=0}|^2$, the momentum scattering rate limited by it is $\propto k$.

\subsection{Spin polarization induced by the electric field}
\label{ana1}
Retaining the lowest three orders of ${\bf S}_k^l$ in Eq.~(\ref{ksbe-ana}), i.e., those with $l=0$
and $\pm 1$, one can obtain the steady-state solution as
\begin{align}
%&{\bf
 % S}_k^0(+\infty)=(0,0,S_{kz}^0(0))=(0,0,\frac{\hbar\Delta_k^0}{2}),\label{s1}\\
&{\bf S}_k^0(+\infty)=(0,0,\hbar\Delta_k^0/2),\label{s1}\\
&S_{kx}^{\pm 1}(+\infty)=\frac{\mp ieE}{8 v_fk}(\Delta_k^0k^{-1}+\Delta_k^1),\\
&S_{ky}^{\pm 1}(+\infty)=\frac{\pm ieE}{16(v_fk)^2\tau_k}(\Delta_k^0k^{-1}+\Delta_k^1),\\
&S_{kz}^{\pm 1}(+\infty)=\frac{eE\tau_k}{4}\left[\Delta_k^1+\frac{\Delta_k^0k^{-1}+\Delta_k^1}{4(v_fk\tau_k)^2}\right],\label{s2}
\end{align}
where $\tau_k\equiv\tau_{1k}^1=\tau_{2k}^1$. In the weak scattering limit with
$v_f  \langle k\tau_k\rangle \gg 1$, ${\bf S}_{k}^{\pm 1}(+\infty)$ can be
approximated as ${\bf S}_k^{\pm
  1}(+\infty)\approx eE\tau_k\Delta_k^1\hat{\bf z}/4$. The steady-state solution $S_{kz}^0(+\infty)=S_{kz}^0(0)$ means that
the electron density in each band keeps unchanged under the small
electric field in the framework of this analytical study.

Now we focus on the steady-state spin polarization in the collinear spin
space. According to Eq.~(\ref{sd}),  the spin polarization reads ${\tilde {\bf
    S}}=\frac{1}{2\pi}\int_0^\infty dk k{\tilde{\bf S}}_k^0$ with
\begin{align}
{\tilde{\bf S}}_k^{0}=(-{\rm Im} S_{kz}^1+{\rm Re} S_{ky}^1,-{\rm
  Im}S_{ky}^1-{\rm Re}S_{kz}^1,-S_{kx}^0).
\label{tilde_sk0}
\end{align}
With the solution of ${\bf S}_k^{0,1}(+\infty)$ obtained above, we have
\begin{align}
{\tilde{\bf S}}_k^{0}(+\infty)\approx -eE\tau_k\Delta_k^1{\hat{\bf y}}/4. 
\label{sk0}
\end{align}
Therefore, the total spin polarization ${\tilde {\bf S}}(+\infty)\approx
eE\langle \tau_k\rangle\ln(2+e^{\beta\mu}+e^{-\beta\mu})\hat{\bf
  y}/(8\pi\beta\hbar v_f)$ where
$\langle\tau_k\rangle\equiv\int_0^{+\infty}dkk\Delta_k^1\tau_k/\int_0^{+\infty}dkk\Delta_k^1$. For
the intrinsic nondegenerate case with $\mu=0$, ${\tilde {\bf 
    S}}(+\infty)\approx eE\langle \tau_k\rangle\ln
  4\hat{\bf y}/(8\pi\beta\hbar v_f)$, and for the $n$-type degenerate one, ${ \tilde {\bf
    S}}(+\infty)\approx eE\tau_{k_f} k_f\hat{\bf y}/(8\pi)$
where $k_f$ is the Fermi momentum. The spin polarization obtained for the degenerate case is consistent with
that presented in Ref.~\onlinecite{culcer860} 
where the valence band is fully occupied. These results indicate that under
the small electric field, a transverse spin polarization is
induced due to the Rashba spin-orbit coupling [Eq.~(\ref{h0})], with the magnitude
proportional to the electric field and momentum scattering time.

We further study the momentum scattering time in the intrinsic nondegenerate and
$n$-type degenerate cases respectively, with the help of
Eqs.~(\ref{st1})-(\ref{st2}). For the intrinsic case, the mean
momentum scattering times limited by the electron-impurity and electron-phonon
scatterings are $\langle\tau_k^{\rm ei}\rangle\approx
8\beta\hbar n_e/(\pi r_s^2n_i\ln 4)$ and $\langle\tau_k^{\rm
  ep}\rangle\approx 4\beta\hbar^3v_f^2/[(2N_0+1)|g_0|^2\ln 4]$,
respectively. For the $n$-type case, at the Fermi level
$\tau_{k_f}^{\rm  ei}\approx k_f/[\pi r_s^2v_fn_iI(r_s)]$ and 
$\tau_{k_f}^{\rm ep}\approx 4\hbar^2v_f/[(2N_0+1)|g_0|^2k_f]$. Here
$I(r_s)\equiv\int_0^{2\pi}d\theta \sin^2\theta/(2\sin\frac{\theta}{2}+r_s)^2$ 
and $r_s=e^2/(4\pi\epsilon_0\kappa_0\hbar v_f)$, with
$\epsilon_0$ being the static dielectric constant [when
$\kappa_0=100$, one has $r_s\approx 0.044$ and $I(r_s)\approx 2.6$]. To calculate the momentum scattering time
limited by the electron-phonon scattering, we have neglected the weak
momentum dependence of $|g_{\bf q}|^2$ by approximating it as
$|g_0|^2$.
 
Consequently, the induced spin polarization limited by the electron-impurity scattering
is ${\tilde {\bf S}}^{\rm ei}(+\infty)\approx eEn_e\hat{\bf y}/(\pi^2 r_s^2v_fn_i)$
for the intrinsic nondegenerate case and ${\tilde {\bf S}}^{\rm ei}(+\infty)\approx eEn_e\hat{\bf 
  y}/[2\pi v_fr_s^2I(r_s)n_i]$ for the $n$-type degenerate one. However, the one
limited by the electron-phonon scattering is ${\tilde
  {\bf S}}^{\rm ep}(+\infty)\approx eE\hbar^2v_f\hat{\bf y}/[2\pi |g_0|^2(2N_0+1)]$, in spite of the electron density.
Considering both scatterings, the induced spin polarization along the $y$-axis reads
\begin{align}\nonumber
  {\tilde S}_y(+\infty) &\approx [1/{\tilde S}_y^{\rm ei}(+\infty)+1/{\tilde S}_y^{\rm
      ep}(+\infty)]^{-1}\\
  & =\frac{E\cdot 10^{12}~\mbox{cm/kV}}{\alpha_1 (2N_0+1)+\alpha_2 n_i/n_e}\hbar\cdot\mbox{cm}^{-2}, 
\label{ana-eq}
\end{align}
with $\alpha_1\approx 13$ and $\alpha_2\approx 0.6$ (1) for
the intrinsic nondegenerate ($n$-type degenerate) case. It is interesting to see
that, when the electron-phonon scattering dominates, 
  the induced spin polarization is only sensitive to the
  temperature. Especially, at high temperature $T\gg
  \hbar\omega_0/k_B\approx 90$~K, one has ${\tilde
    S}_y(+\infty)\propto T^{-1}$. Due to the large relative static dielectric constant, only when
  the impurity density is high enough (e.g., with $n_i\gtrsim 10 n_e$)
  and the temperature is low, the effect of electron-impurity
  scattering can be comparable to that of the electron-phonon
  scattering and then leads to the impurity/electron density dependence of spin polarization.
 
\subsection{Spin relaxation}
We then start from the steady state obtained previously to study the spin
relaxation with the electric field turned off. From the initial state ${\bf
  S}^{0,1}_k(0)=S_{kz}^{0,1}(0){\bf\hat z}$, one can solve the
temporal evolution of ${\bf S}_k^{0,1}(t)$ by Eq.~(\ref{ksbe-ana}) after
the terms proportional to $E$ have been removed. Then 
according to Eq.~(\ref{tilde_sk0}), one obtains $\tilde{\bf S}_k^0(t)={\tilde
  S}_{ky}^0(t)\hat{\bf y}$ with 
\begin{align}
{\tilde S}_{ky}^0(t)={\tilde
  S}_{ky}^0(0)e^{-t/\tau_k}\left[1-c_k^{-2}(1-c_k\sin\frac{c_kt}{\tau_k}-\cos\frac{c_kt}{\tau_k})\right].
\label{sr}
\end{align}
Here ${\tilde S}_{ky}^0(0)=-S_{kz}^1(0)$ and $c_k=\sqrt{(2v_fk\tau_k)^2-1}$. In
the weak scattering limit with $c_k\gg 1$,
${\tilde S}^0_{ky}(t)\approx{\tilde S}^0_{ky}(0)e^{-t/\tau_k}$, indicating
that the spin polarization relaxes in the time scale of momentum
scattering. This feature is in agreement with that given by Schwab
{\it et al.}\cite{schwab67004} and Burkov and Hawthorn.\cite{burkov066802}

\section{Numerical results}
\label{num}
The analytical study in the previous section only applies to the small electric
field. In order to take into account the large electric field, as 
well as all the scatterings explicitly, we carry
out the numerical calculation based on the KSBEs. In the
calculation, we first apply the electric field along the $x$-axis to
study the charge and spin transport, and then turn off the electric
field after reaching the steady state to look into the cooling of the hot carriers and the relaxation
of the previously induced spin polarization. We consider both the intrinsic
nondegenerate and $n$-doped degenerate cases, starting from the initial equilibrium state with $\mu=0$ and a
given $n_e(0)$ respectively. In our study $n_e(0)$ is
  chosen to be of the order of 10$^{11}$~cm$^{-2}$ or even smaller, to
  avoid entering the bulk states in the presence of high electric field and
  also ensure the validity of Hamiltonian $H_0$ [Eq.~(\ref{h0})] without involving the
  terms square or cubic in momentum.\cite{culcer155457,fu103} We also do not consider the intrinsic case 
  under low temperature as for such case the carrier density is quite low and the
  fluctuation (e.g., the effect of puddles\cite{kim459}), which is beyond the
  scope of this work, becomes important. 

\subsection{Redistribution of electrons between two bands}  
 
In semiconductors with a large band gap, the electron and hole densities in
  the conduction and valence bands remain unchanged under the static
electric field. However, here on the surface of topological insulator, due to
the spin mixing of the two bands, the static electric field leads to inter-band
precession,\cite{balev165432} and also, due to the zero band gap, the inter-band electron-phonon
scattering is easy to take place. Therefore, electrons can be transferred from the
valence band to the conduction one under the influence of the electric
field, with the difference in the densities of
electrons and holes in the two bands, $n_e-n_h$, keeping constant due
to the particle conservation. This redistribution of electrons between two
  bands under the electric field has also been revealed in a similar
  system, the gapless graphene.\cite{balev165432} In principle, 
  the Auger process of the Coulomb scattering, during which one and only one of the
  two scattered electrons transfers between two
  bands,\cite{winzer4839,winzer241404} may also lead to the redistribution of electrons between two bands under 
  the electric field. However, this process is actually forbidden when the dynamic screening under random phase 
approximation (RPA)\cite{hwang205418,ramezanali214015,culcer155457} (refer to Appendix~\ref{ap1}) is
adopted.\cite{yzhou,bysun125413} The redistribution of electrons in
the two bands can not be revealed by the analytical study presented in
Sec.~\ref{ana}, which fails to  incorporate the 
inter-band precession [due to the replacement of $\rho_{\bf k}(t)$ by
$\rho_{\bf k}(0)$ in terms proportional to $E$] as well as the inter-band scattering (due to the elastic scattering
approximation). However, this process can be revealed by fully solving the KSBEs, as stated in the following.

\begin{figure}[hbt]
  {\includegraphics[width=8.8cm]{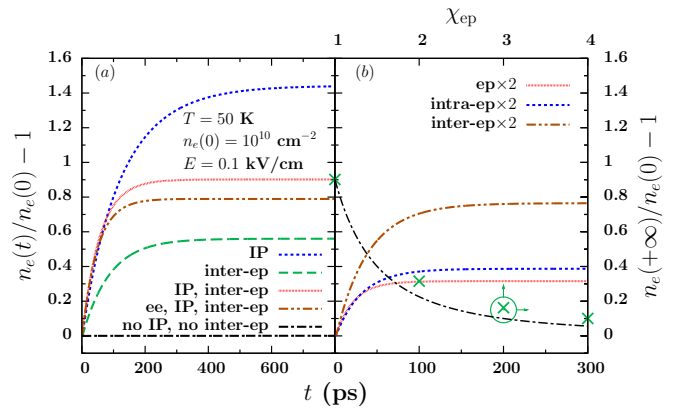}}
  \caption{(Color online) Temporal evolution of $n_e(t)/n_e(0)-1$ with
    $E=0.1$~kV/cm, $T=50$~K and $n_e(0)=10^{10}$~cm$^{-2}$. During calculation, the 
    electron-impurity scattering is not included while the intra-band 
    electron-phonon scattering (labeled as intra-ep) is always present. (a): The chain
    curve is calculated without the inter-band precession (labeled as IP) and inter-band electron-phonon
    scattering (labeled as inter-ep), the dashed (dotted) one with
    only the inter-band electron-phonon scattering (inter-band precession),
    while the solid one with both the inter-band precession and inter-band
    electron-phonon scattering. The double-dotted chain curve is calculated with
    the inter-band precession, inter-band electron-phonon scattering and
    also the electron-electron Coulomb scattering (labeled as ee). (b): Temporal
    evolution of $n_e(t)/n_e(0)-1$ with the electron-phonon  
    scattering artificially strengthened. The inter-band precession is always included. The
    solid curve is calculated with both the intra- and inter-band
    electron-phonon scatterings strengthened by 
    a factor $\chi_{\rm ep}=2$, while the dotted (double-dotted chain) one with only the
    intra-band (inter-band) electron-phonon scattering strengthened by a factor
    2. The crosses (with scales on the top and right-hand side of the 
    frame) are the relative variation of $n_e$ reached in the steady state,
    $n_e(+\infty)/n_e(0)-1$, against the modulation factor $\chi_{\rm
      ep}$ ranging from 1 to 4. The chain curve (with scales on the top and right-hand side of the
    frame) is the function $[n_e(+\infty)/n_e(0)-1]|_{\chi_{\rm ep}=1}/\chi_{\rm  ep}^{2}$.} 
  \label{figzw1} 
\end{figure}  
  
We take an $n$-type degenerate case with $T=50$~K and
$n_e(0)=10^{10}$~cm$^{-2}$ (the corresponding Fermi energy is
$\varepsilon_f\approx 12$~meV and the Fermi momentum is $k_f\approx
0.035$~nm$^{-1}$) as an example to show the temporal evolution of the relative
change of electron density in the conduction band, $n_e(t)/n_e(0)-1$, under
electric field $E=0.1$~kV/cm. It is seen in Fig.~\ref{figzw1}(a) that when
both the inter-band precession and inter-band electron-phonon scattering are
excluded, $n_e(t)$ remains almost unchanged (chain 
curve). However, once the inter-band precession is included, electrons can be effectively
transferred from the lower band to the higher one (dotted curve). The scenario
is that, the electrons are precessed from the states below the Dirac point to the ones above and then driven away to 
higher-energy states by the electric field. The transfer of electrons can
also be alternatively realized with the assistance of phonons (dashed 
curve). Around the Dirac point (with $k< k_0\equiv\omega_0/v_f\approx
0.022$~nm$^{-1}$), when electrons in the conduction band are
driven away to higher-energy states, electrons in the valence band tend to enter
the conduction band by {\em absorbing} phonons. However, the above two effects are
not definitely superimposed when both of them are present. When the inter-band
precession transfers electrons between two bands effectively, i.e., leads to
substantial nonequilibrium between the two bands (such as the case presented
here), electrons in the conduction band tend to fall back to the valence band by
{\em emitting} phonons. Therefore, when both the inter-band scattering and
inter-band precession are included, the steady-state 
electron density in the conduction band decreases compared to the case with
only the inter-band precession (compare the solid and dotted curves).

As pointed out above, the inter-band precession and inter-band scattering open
channels for electron transfer between the two bands. However, the steady state
of the two-band system is determined by the balance between the rates of energy
gain from electric field and the energy loss to phonons. The former is
determined by $\partial_t\varepsilon_i(t)=E{\tilde
  j}_x(t)\propto E{\tilde S}_y(t)$ [refer to Eq.~(\ref{current})]. If the steady state is not far
away from the equilibrium, one approximately has $\partial_t\varepsilon_i(+\infty)\propto
E^2|g_0|^{-2}$ in the presence of electron-phonon
scattering only. The dominant channel of the energy loss is the intra-band
electron-phonon scattering. That is because the inter-band electron-phonon
scattering is limited to a small finite region $k<k_0$ in momentum space, and
also, with such small momentum, the scattering is weak as the rate is $\propto
k$. A rough estimation gives that the rate of energy loss due to the
intra-band electron-phonon scattering for the degenerate case with both low
temperature and electron density is $\partial_t\varepsilon_o(t)\propto
|g_0|^2[n_e(t)-n_e(0)]$ (refer to Appendix~\ref{ap2}). Therefore in the
steady state not far away from the equilibrium, one has  
\begin{align}
  n_e(+\infty)-n_e(0)\propto E^2|g_0|^{-4}, 
\label{nv}
\end{align} 
required by  $\partial_t\varepsilon_i(+\infty)=\partial_t\varepsilon_o(+\infty)$. 
 
Based on Eq.~(\ref{nv}), it is found that the strengthening of the
electron-phonon scattering, especially the intra-band part, 
leads to the decrease in $n_e(+\infty)/n_e(0)-1$. In Fig.~\ref{figzw1}(b) we
numerically verify this by performing similar calculation as in (a) but with the
electron-phonon scattering artificially strengthened. It is shown by the solid curve that when the total 
electron-phonon scattering is strengthened by a factor $\chi_{\rm ep}=2$ (solid
curve), $n_e(+\infty)/n_e(0)-1$ substantially decreases compared to the genuine case [solid curve in
(a)]. When only either the intra- or inter-band electron-phonon scattering is
strengthened (dotted and double-dotted chain curves, respectively),
$n_e(+\infty)/n_e(0)-1$ also decreases. However, the decrease is not obvious
when only the inter-band electron-phonon scattering is strengthened (double-dotted chain
curve), indicating that the contribution to the energy loss by the inter-band
electron-phonon scattering is indeed relatively weak. 
%Besides, we also show the case in the absence of inter-band precession but with
%the inter-band electron-phonon scattering strengthend by a factor 2 (dashed
%curve). The comparison between the dashed curves in (a) and (b) indicates that
%the contribution to energy loss by the inter-band electron-phonon scattering is
%indeed marginal. Nevertheless, the strengthened inter-band scattering can
%decrease the time for the system to reach the steady state.
We further show the steady-state value, $n_e(+\infty)/n_e(0)-1$,
against $\chi_{\rm ep}$ ranging from 1 to 4 by the crosses (the scales are on
the top and right-hand side of the frame). It is seen that approximately
$n_e(+\infty)/n_e(0)-1\propto\chi_{\rm ep}^{-2}$,
satisfying Eq.~(\ref{nv}) [as a
guide to the eye, a function proportional to $\chi_{\rm ep}^{-2}$,
$[n_e(+\infty)/n_e(0)-1]|_{\chi_{\rm ep}=1}/\chi_{\rm ep}^{2}$, is plotted by the
chain curve in (b)].

At last we briefly address the effect of electron-impurity and electron-electron
Coulomb scatterings on the redistribution of electrons. Although neither of them
leads to energy loss directly, both of them can limit the charge current and
hence the energy injection rate of the electric field. The effect of impurities
on charge current is apparent, as also explicitly indicated here by the
combination of Eqs.~(\ref{ana-eq}) and (\ref{current}). The Coulomb scattering 
is usually deemed to preserve the charge current. This is indeed the case in
semiconductors with {\em parabolic} energy spectrum where both momentum and
current are conserved during the Coulomb scattering. However, here with {\em linear}
energy spectrum, only the momentum but not the current is conserved during the 
Coulomb scattering. This particular feature also exists in
graphene with linear dispersion as well.\cite{kashuba085415}
In fact, the effect of the Coulomb scattering can be conjectured based on our analytical
study, which indicates that ${\tilde j}_x\propto{\tilde S}_y\propto
\tau_k$ [Eqs.~(\ref{current}) and (\ref{sk0})]. With the addition of the Coulomb scattering, the momentum scattering time 
$\tau_k$ is reduced and hence the charge current decreases. Consequently,
with the inclusion of
the electron-impurity and/or electron-electron Coulomb scattering,
 the current and hence the energy injection 
rate of electric field decreases, resulting in less obvious redistribution of
electrons. Nevertheless, due to the large relative static dielectric constant,
the effect of the electron-impurity and electron-electron Coulomb scatterings is expected to be weak except when the
temperature is low. In Fig.~\ref{figzw1}(a), we add the double-dotted
chain curve calculated with the Coulomb scattering included. In the
absence of impurities and under low temperature, the contribution of the Coulomb scattering is visible, with which
the variation in $n_e$ decreases compared to the Coulomb scattering-free case
(solid curve there).

\subsection{Charge and spin transport} 
In the following, we systematically investigate the charge and spin transport
under the electric field, first in the low electric field regime and then the large one. 

\subsubsection{Low electric field regime}
\label{lf}
\begin{figure}[hbt]
  {\includegraphics[width=8.7cm]{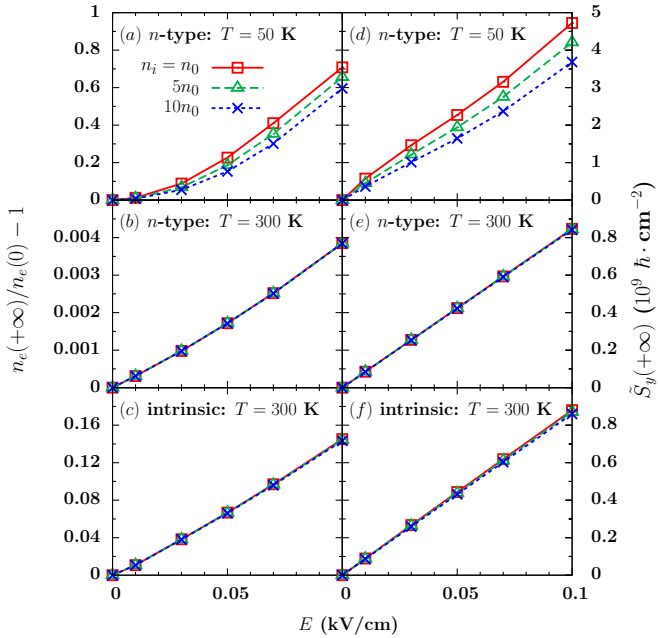}}
  \caption{(Color online) (a)-(c): The relative change of electron density
   in the conduction band, $n_e(+\infty)/n_e(0)-1$, against the electric field
    $E$ for cases (I)-(III), respectively, with different impurity
    densities. $n_0=10^{10}$~cm$^{-2}$. (d)-(f): The induced spin polarization
    ${\tilde S}_y(+\infty)$ against the electric field $E$ for cases (I)-(III)
    with different impurity densities (the scale is on the right-hand side of the frame).}
  \label{figzw2}
\end{figure}

We first focus on the low electric field regime with $E\le
0.1$~kV/cm and compare the numerical results with the analytical
study in Sec.~\ref{ana}. We consider three cases, (I) $n$-doped case with
$T=50$~K and $n_e(0)=10^{10}$~cm$^{-2}$, (II) $n$-doped case with $T=300$~K and $n_e(0)=5\times
10^{11}$~cm$^{-2}$ and (III) intrinsic case with
$T=300$~K and $\mu=0$ [correspondingly $n_e(0)\approx 0.8\times
  10^{11}$~cm$^{-2}$]. Corresponding to these three cases, in
  Figs.~\ref{figzw2}(a)-(c), we show the dependence of $n_e(+\infty)/n_e(0)-1$ 
  on the electric field with different impurity  
densities (note that the $y$-axes are in different scales in the three figures). For case (I) with low temperature 
$T=50$~K, $n_e$ increases obviously with the electric field, due to the small
initial electron density in the conduction band $n_e(0)$ as well as the weak electron-phonon
scattering. Moreover, $n_e$ increases with $E$ faster with a smaller impurity density as the
electron-impurity scattering weakens. Besides, it is shown that when $E$ is
  small ($E<0.05$~kV/cm), $n_e(+\infty)/n_e(0)-1$ roughly exhibits a square
  dependence on $E$, consistent with Eq.~(\ref{nv}). For case (II), which is highly degenerate 
and with strong electron-phonon scattering due to both high temperature and
electron density, the electron density remains almost  
unchanged in the low electric field regime under investigation (the
relative variation is of the order of $10^{-3}$). Nevertheless, for
the intrinsic case (III), it becomes relatively easier for electrons to
transfer due to the less occupancy of electrons in the 
conduction band, when compared to case (II). Finally,
Figs.~\ref{figzw2}(b) and (c) show that the effect of impurities
is negligible at $T=300$~K.

The induced spin polarizations in the steady state ${\tilde S}_y(+\infty)$ against the electric field $E$
for the three cases (I)-(III) are plotted in
Figs.~\ref{figzw2}(d)-(f), respectively. It is shown that ${\tilde S}_y(+\infty)$ increases
with $E$ much faster at 50~K than those at 300~K, mainly due to the weaker electron-phonon
scattering. For case (I) with $T=50$~K, the
electron-impurity scattering is important and $n_e$ increases with $E$
effectively [refer to Fig.~\ref{figzw2}(a)], therefore the rate of the increase in ${\tilde S}_y(+\infty)$ with
$E$ shows impurity density dependence and also a slight electric field
dependence [refer to Eq.~(\ref{ana-eq})]. However, for  cases (II) and (III)
with $T=300$~K and hence the strong electron-phonon scattering, ${\tilde
  S}_y(+\infty)$ increases with $E$ linearly, with almost the identical rate
under different impurity and electron densities.  
This is consistent with the analytical study: when the electron-phonon
scattering dominates, the spin polarization increases with the electric field
linearly in a rate solely determined by temperature [refer to Eq.~(\ref{ana-eq}) and the discussion there].

\begin{figure}[hbt]
  {\hspace{-0.5 cm}\includegraphics[width=6.5cm]{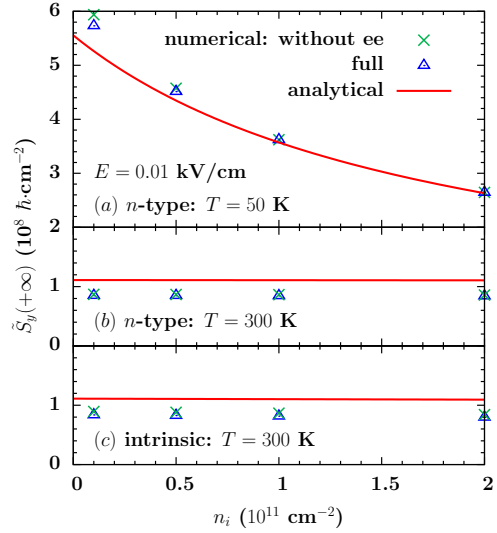}}
  \caption{(Color online) Electric-field--induced spin polarization ${\tilde
      S}_y(+\infty)$ in the steady state against impurity
    density $n_i$ for cases (I)-(III), in (a)-(c) respectively. The triangles are obtained by numerically solving the full 
    KSBEs, while the crosses are calculated without the electron-electron
    Coulomb scattering (labeled as ee). The solid curves are from the analytical result [Eq.~(\ref{ana-eq})]. }
  \label{figzw3} 
\end{figure} 

In the following, we compare the induced steady-state
 spin polarization numerically obtained
to the analytical one at $E=0.01$~kV/cm, with which all
the three cases (I)-(III) are close to the equilibrium. In Fig.~\ref{figzw3}, the 
impurity density dependence of the spin polarization from the numerical calculation
with all the scatterings explicitly included is shown by triangles for the three
cases. We also plot the numerical results without the Coulomb scattering 
by crosses. It is shown that the effect of the Coulomb scattering can only be visible when the impurity density approaches
zero and the temperature is low. For comparison, the analytical calculation based on Eq.~(\ref{ana-eq}) is  
plotted by the solid curves in the figure. One finds that the analytical
formula primarily captures the numerical results. We point out that at 300~K where the
elastic scattering approximation for electron-phonon scattering is
more reasonable ($k_BT\gg \hbar\omega_0$), the best fit to the numerical results by Eq.~(\ref{ana-eq}) requires
$\alpha_1=17$. Finally, Figs.~\ref{figzw3}(b) and (c) indicate more
clearly that at room temperature the electron-impurity scattering is negligible even when $n_i$ reaches $2\times
10^{11}$~cm$^{-2}$. In fact, according to the analytical study in
Sec.~\ref{ana1}, the momentum scattering times are estimated to
be $\tau_{k_f}^{\rm ei}=4.5$~ps and  $\tau_{k_f}^{\rm ep}=2.5$~ps for case (I),
$\tau_{k_f}^{\rm ei}=31.8$~ps and $\tau_{k_f}^{\rm ep}=0.07$~ps for case (II),
and $\langle \tau_{k}^{\rm ei}\rangle=19.7$~ps and $\langle\tau_{k}^{\rm
  ep}\rangle=0.17$~ps for case (III) (in the estimation the impurity density is
set as $n_i=10^{11}$~cm$^{-2}$). These values quantitatively support the
  dominance of the electron-phonon scattering at high temperature.

\subsubsection{Large electric field regime}
After investigating the low field regime and comparing
with the analytical study, we proceed with the large electric field
regime ($E$ is upto 7~kV/cm). Both the intrinsic nondegenerate and $n$-type
degenerate cases under 150~K and 300~K are considered. For the  
degenerate case, we set $n_e(0)=5\times 10^{11}$~cm$^{-2}$. %Therefore,
%$n_e-n_h$, as a constant, is zero for the intrinsic case and about $5\times
%10^{11}$~cm$^{-2}$ for the $n$-type degenerate one. 
The impurity density is fixed at $n_i=10^{11}$~cm$^{-2}$, with which the 
electron-impurity scattering is in fact negligible under the temperature 
investigated. 

\begin{figure}[hbt]
  {\includegraphics[width=6.5cm]{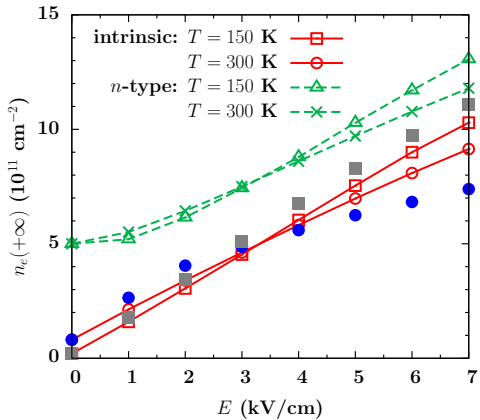}}
  \caption{(Color online) Electric field dependence of the steady-state electron density in the
    conduction band $n_e(+\infty)$ for the intrinsic nondegenerate (solid curves) and $n$-type degenerate
    (dashed curves) cases under different temperatures. The dots are
    calculated for the intrinsic case at $T=300$~K by setting $\lambda_1=0$ and 
    $\lambda_2=20$~eV$\cdot$nm$^2$ in the electron-phonon scattering matrix
    element [Eq.~(\ref{ep-me})]. The closed squares are calculated without the electron-electron Coulomb scattering
    for the intrinsic case at 150~K.}
  \label{figzw4} 
\end{figure}

We first present the electric field dependence of the steady-state electron density
in the conduction band, $n_e(+\infty)$, for the intrinsic and $n$-type cases
at different temperatures. Figure~\ref{figzw4} indicates that $n_e(+\infty)$ increases with $E$
almost linearly for both cases, except in the low electric field
region $E<2$~kV/cm for the degenerate situation where the Pauli
blocking is important. In fact, a nearly linear increase of
electron density in conduction band with electric field is also
observed in intrinsic graphene when $E>0.01$~kV/cm.\cite{balev165432} It is also
noted that here in the nearly linear regime, the rate
of the increase in $n_e(+\infty)$ with $E$ is determined by temperature and is
insensitive to the electron density. It is believed that this linear relation as
well as the temperature dependent increasing rate are attributed to the dominant
electron-phonon scattering with almost constant scattering matrix
element. As a comparison, we recalculate the intrinsic case at 300~K by
artificially setting $\lambda_1=0$ and $\lambda_2=20$~eV$\cdot$nm$^2$ in the
electron-phonon scattering matrix element [Eq.~(\ref{ep-me})]. The result is shown in the figure by
dots. It is seen that it deviates from the linear relation obviously. Finally,
we also present the calculation without the electron-electron Coulomb scattering
for the intrinsic case at 150~K by the closed squares. The comparison between
the closed and open squares indicates that the influence of the Coulomb
scattering is marginal when the electric field is low and relatively effective
when the electric field is high.

\begin{figure}[hbt]
  {\includegraphics[width=6.5cm]{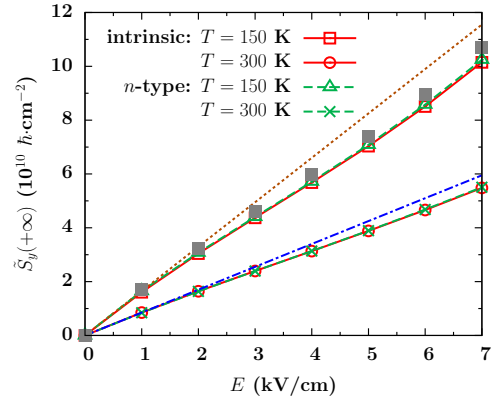}}
  \caption{(Color online) Electric field dependence of induced spin polarization
    in the steady state for the intrinsic (solid curves) and $n$-type degenerate 
    (dashed curves) cases at different temperatures. The dotted and chain
    curves are obtained from Eq.~(\ref{ana-eq}) with $\alpha_1=17$ at
    150~K and 300~K, respectively. The closed squares are calculated without the electron-electron Coulomb scattering
    for the intrinsic case at 150~K.}
  \label{figzw5}  
\end{figure}
We then look into the steady-state spin polarization ${\tilde S}_y(+\infty)$ induced by the electric field. 
In Fig.~\ref{figzw5} the electric field dependence of ${\tilde S}_y(+\infty)$
for the intrinsic and $n$-type cases at different temperatures is
plotted. Strikingly, in this whole large electric 
field region with the electron-phonon scattering being dominant, ${\tilde
  S}_y(+\infty)$ is solely determined by temperature, as previously revealed in
the low electric field regime. Moreover, with the same electric field $E$,
${\tilde S}_y(+\infty)$ at 150~K is about 2 times as large as that at
300~K, approximately satisfying the analytical relation ${\tilde S}_y(+\infty)\propto 
T^{-1}$ at high temperature. In the figure we also plot the ${\tilde S}_y(+\infty)$-$E$ relation 
given by Eq.~(\ref{ana-eq}) (with modified $\alpha_1=17$) at 150~K (dotted curve)
and 300~K (chain curve), respectively. It is seen that the
analytical formula for ${\tilde S}_y(+\infty)$ can approximately apply up to $E_c=2$~kV/cm
(4~kV/cm) at 150~K (300~K). When $E$ exceeds $E_c$, the electron heating
becomes important. Therefore, with the occupation of 
larger-momentum states, the electron-phonon scattering is
strengthened. At the same time, the momentum dependence of the electron-phonon
scattering matrix element [arising from the term $\lambda_2q$ in Eq.~(\ref{ep-me})]
also plays a role and further enhances the
electron-phonon scattering. As a result, ${\tilde S}_y(+\infty)$ tends to
decrease and hence deviates from the linear relation against  
$E$. Finally, the result obtained without the electron-electron Coulomb scattering
for the intrinsic case at 150~K is also plotted by the closed squares. Again,
the effect of the Coulomb scattering on spin polarization is also shown to be marginal,
especially in the low electric field regime. 

\begin{figure}[hbt]
  {\includegraphics[width=6.5cm]{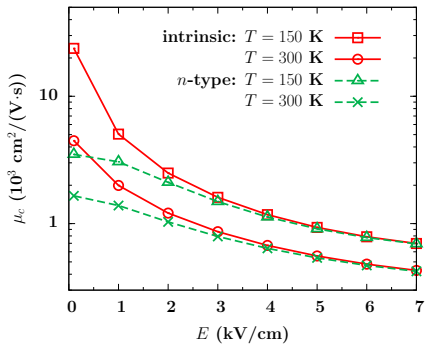}}
  \caption{(Color online) Electric field dependence of mobility for the
    electron-hole system in the intrinsic (solid curves) and $n$-type (dashed curves) cases at different temperatures.} 
  \label{figzw6} 
\end{figure}

We now turn to study the mobility of the two-band system. According to
Eq.~(\ref{current}), the steady-state charge current under the electric field is
immediately obtained as ${\tilde j}_x(+\infty)=2ev_f{\tilde S}_y(+\infty)/\hbar$.
Therefore, the steady-state mobility of the electron-hole system  
can be determined by $\mu_c={\tilde j}_x(+\infty)/[eE(n_e(+\infty)+n_h(+\infty))]$. In Fig.~\ref{figzw6} we
plot $\mu_c$ against $E$ for the intrinsic and $n$-type cases at different
temperatures. It is shown that $\mu_c$ is around the order of
$10^3$~cm$^2$/(V$\cdot$s), in consistence with the experimental
data.\cite{wei201402,kim459,analytis960} $\mu_c$ decreases with $E$ as electrons
and holes are heated to occupy large-momentum states where strong electron-phonon
scattering takes place. Moreover, with the increase of temperature from 150 to 300~K,
$\mu_c$ decreases as well. In fact, we roughly have 
$\mu_c\approx ({\mu_c^0}^{-1}+\gamma E)^{-1}$ as both $n_e(+\infty)+n_h(+\infty)$ and ${\tilde j}_x(+\infty)$ are close to
linear functions of $E$. Here $\gamma$ is determined by the increasing rates of
$n_e(+\infty)$ and ${\tilde j}_x(+\infty)$ with $E$, both of which are only sensitive
to temperature when $E$ is large (as indicated by Figs.~\ref{figzw4} and
\ref{figzw5}). Therefore, in the large electric field regime, $\mu_c\approx  
\gamma^{-1}E^{-1}$, with $\gamma$ determined by the temperature. This feature is
manifested in Fig.~\ref{figzw6} when $E\gtrsim 3$~kV/cm. 

\subsection{Effect of the Coulomb scattering}% and carrier heating}

\begin{figure}[hbt]
  {\includegraphics[width=7cm]{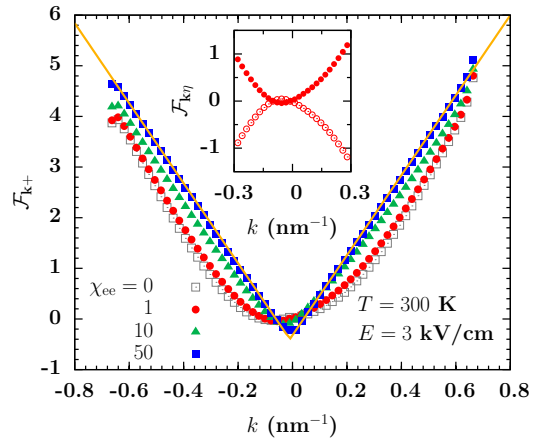}}
  \caption{(Color online) Function ${\cal F}_{{\bf k}+}\equiv \log(1/f_{{\bf k}+}-1)$
    against momentum $k$ along the $x$-axis in the steady state, evolving from
    the intrinsic case with $T=300$~K and $E=3$~kV/cm. The dots are obtained
    from the calculation for the genuine case, while the 
    open squares, triangles and closed squares are from the calculation with the Coulomb scattering term
    multiplied by $\chi_{\rm ee}=0$, 10 and 50, respectively. The solid line is
    plotted as a guide to the eye by fitting the closed squares with
    hot-electron temperature $T_e\approx 488$~K. Inset: Function
    ${\cal F}_{{\bf k}\eta}\equiv \log(1/f_{{\bf k}\eta}-1)$ [closed (open) circles are for
    electrons in conduction (valence) band with $\eta=+$ ($-$)] against
    momentum $k$ along the $x$-axis in a small momentum scale in the steady
    state for the genuine case.}
  \label{figzw7} 
\end{figure}

Now we turn to see the effect of the Coulomb scattering on the steady state established
in the presence of electric field. It has been revealed in the previous section that due to the large relative static 
dielectric constant, the influence of the Coulomb scattering is marginal in the presence of
electron-phonon and electron-impurity scatterings. With the weak Coulomb scattering, electrons
in two bands fail to reach the drifted Fermi distribution with a
unified hot-electron temperature in the steady state. To reveal this, we take the intrinsic case with $T=300$~K and
$E=3$~kV/cm as an example to plot the dependence of function ${\cal F}_{{\bf k}+}\equiv \log(1/f_{{\bf k}+}-1)$
on $k$ along the $x$-axis in Fig.~\ref{figzw7}, with different Coulomb
scattering strengths adjusted artificially.\cite{yzhou} The comparison among these sets of
data indicates that only when the Coulomb scattering
is strong enough (e.g., with the Coulomb scattering term rescaled by a factor $\chi_{\rm ee}=50$), in the steady state the
drifted Fermi distribution $f_{{\bf k}\pm}=1/\{\exp[(\hbar v_f|{\bf
  k}-k_x^0|-\mu_\pm)/(k_BT_e)]+1\}$ is reached. Here $T_e$ is the unified
hot-electron temperature and $k_x^0$ is the shift of the momentum
center limited by the scattering. Both can be obtained from the slope of the 
wings of the ``V'' shape and the position of the valley,
respectively (refer to the solid curve as a guide to the eye). $\mu_\pm$ are
chemical potentials for the two bands. For the intrinsic case as presented here, $\mu_+=-\mu_-$, 
due to the symmetry between the two bands. In the inset of the figure we plot both
${\cal F}_{{\bf k}+}$ (closed circles) and ${\cal F}_{{\bf k}-}$ (open circles)
for the genuine case (with $\chi_{\rm ee}=1$) in a small momentum
scale. Although the drifted Fermi distribution is not established, the
symmetry between the conduction and valence bands is clearly indicated by the inset. 

\begin{figure}[hbt]
  {\includegraphics[width=6.8cm]{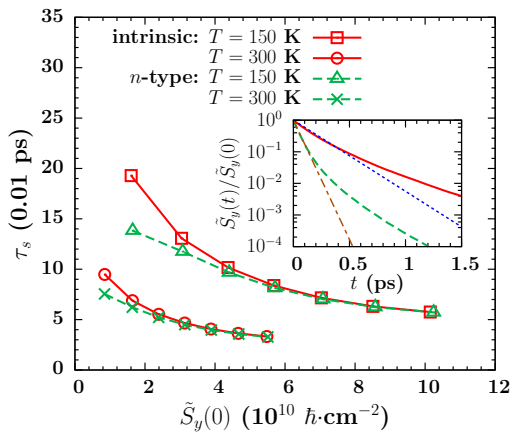}}
  \caption{(Color online) Spin relaxation time $\tau_s$ against the initial spin
    polarization ${\tilde S}_y(0)$ for the intrinsic (solid curves) and $n$-type
    (dashed curves) cases at different temperatures. The temporal evolution of
    ${\tilde S}_y(t)/{\tilde S}_y(0)$ is plotted in the inset for two
    cases, with ${\tilde S}_y(0)$ initialized from the intrinsic situation at
    150~K under electric field 1 (solid curve) and 7~kV/cm (dashed curve),
    respectively. The dotted (chain) curve is the exponential fit to the rapid
    decay in the beginning of the solid (dashed) curve. } 
  \label{figzw8} 
\end{figure}

\subsection{Spin relaxation}
Finally, we start from the steady-state spin polarization reached previously under the electric
field, to study the spin relaxation by turning off the
electric field. The zero point of time, $t=0$, is reset to be the moment at
which the electric field is turned off. Beginning at $t=0$, the heated electrons   
cool down to the original Fermi distribution in the time scale of energy relaxation, which is
as long as $\sim$100-1000~ps, mainly due to the weak inter-band
  electron-phonon scattering.
 However, the spin polarization relaxes in the time scale
of momentum scattering which is quite short.\cite{schwab67004,burkov066802}
According to Eq.~(\ref{sr}), ${\tilde S}_{ky}(t)$ relaxes approximately in the
exponential form when $v_fk\tau_k\gg 1$. The total spin polarization ${\tilde
  S}_y(t)$, containing a summation of ${\tilde S}_{ky}(t)$ over momentum, does not definitely relax in a fine 
exponential form. Our calculation shows that ${\tilde S}_y(t)$ against $t$
decays first very fast to about $0.1{\tilde S}_y(0)$ but then slowly. The
slowly-decaying tail with small magnitude is determined by the low-energy states near the Dirac point with long momentum
relaxation times. As an example, in the inset of Fig.~\ref{figzw8}, we plot the dependence
of ${\tilde S}_y(t)/{\tilde S}_y(0)$ against $t$ for two cases by the solid and dashed
curves. In these two cases ${\tilde S}_y(0)$ are initialized from the intrinsic 
situation at 150~K under electric field 1 (solid curve) and 7~kV/cm (dashed
curve), respectively. To obtain the spin relaxation time $\tau_s$, we fit the rapid
decaying part in the beginning by an exponential function, as illustrated by the dotted and chain
curves in the inset. 

The spin relaxation time $\tau_s$ against the initial spin polarization
${\tilde S}_y(0)$ for the intrinsic and $n$-type cases at different
temperatures is plotted in Fig.~\ref{figzw8}. It is shown that the spin relaxation
time is of the order of 0.01-0.1~ps, mainly limited by the electron-phonon
scattering. In fact, a momentum relaxation time of the similar order is given
from both estimation and experiment by Butch {\sl et al.}.\cite{butch241301} The
decrease of $\tau_s$ with the increase of ${\tilde S}_y(0)$ is due to the fact that the larger
${\tilde S}_y(0)$ is initialized by the higher electric field, with which the
electrons and holes occupy larger-momentum states and 
hence exhibit faster momentum relaxation. Besides, with higher temperature, the
spin relaxation time also decreases as the electron-phonon scattering is strengthened.

\section{Conclusion}
In conclusion, we have studied the charge and spin transport under the influence
of high electric field (up to several
kV/cm) on the surface of topological insulator Bi$_2$Se$_3$, by means
of the KSBEs. We assume that the Fermi level, adjustable by
doping,\cite{analytis960,kulbachinskii15733,wei201402,xia398,hor195208,ren075316}
is located in the bulk gap. Therefore the bulk states are excluded from our
study. With moderate electron and hole densities, the surface state around the
Dirac point can be depicted by the Rashba spin-orbit
coupling. In our study, both the conduction and valence bands of the surface state are considered,
with the inter-band coherence explicitly included. Apart from the driving effect in
each band, due to the spin mixing of conduction and valence bands, the electric
field also leads to inter-band precession. This differs from the semiconductors
with parabolic energy spectrum.

In Bi$_2$Se$_3$, the relative static dielectric constant is as large as
100, indicating weak electron-impurity and electron-electron Coulomb
scatterings. With the weak Coulomb scattering, electrons in the
two bands fail to establish a drifted Fermi distribution with a unified
hot-electron temperature under the driving of the electric field. The electron-surface optical 
phonon scattering dominates in a large temperature region. Moreover, the
electron-phonon scattering matrix element is approximately constant
due to its marginal dependence on momentum. This feature leads to
particular properties of charge and spin transport on the surface of
Bi$_2$Se$_3$. 

Our study reveals that in the presence of driving of the electric
field, both the inter-band precession and inter-band electron-phonon scattering cause
electrons to transfer from the valence band to the conduction
one. Due to the dominant electron-phonon scattering, the variation in
electron density for each band is linear in the electric field when the latter is
high, despite whether the initial state is degenerate or nondegenerate. It is also
found that due to the spin-momentum locking from the Rashba
spin-orbit coupling, a transverse spin polarization is induced by the electric
field, with the magnitude proportional to the momentum scattering 
time. Besides, the spin polarization is linear in the electric field when the latter is
small but deviates from the linear relation when the latter is 
large enough as the electron-phonon scattering is enhanced due to
the heating of electrons. Moreover, a very interesting feature is that at high temperature, the spin polarization is
inversely proportional to the temperature but insensitive 
to the electron density. 

The cooling of hot carriers and the relaxation of spin polarization induced by
the electric field are investigated by turning off the electric field after
reaching the steady state. It is found that the hot carriers cool down in a time
scale of energy relaxation, which is quite long and is of the order of
100-1000~ps. However, due to the spin-momentum locking again, the spin 
polarization relaxes in a time scale of momentum scattering. The spin
polarization is mainly contributed by the states with large momentum. Therefore,
it decays rapidly within the time of the order of 0.01-0.1~ps. Following this
rapid decay, there is a slowly damping tail. This tail is attributed to the low-energy
states near the Dirac point where the momentum scattering is weak.

\begin{acknowledgments}

This work was supported by the National Basic Research 
Program of China under Grant
No.\,2012CB922002 and the Strategic Priority Research Program of the Chinese
Academy of Sciences under Grant No. XDB01000000. One of the authors (PZ) would
like to thank M. Q. Weng for valuable discussions.
\end{acknowledgments}

\begin{appendix}
\section{Scattering term in KSBEs}
\label{ap1}
%The hartree-fock self energy in Eq.~(\ref{ksbe}) is 
%\begin{align}
%\Sigma_{\bf k}(t)=-\sum_{{\bf k}^\prime}V_{{\bf k}-{\bf
%    k}^\prime}^0\Lambda_{{\bf k}{\bf k}^\prime}[\rho_{{\bf
%    k}^\prime}(t)-\frac{1}{2}(1-\sigma_z)]\Lambda_{{\bf k}^\prime{\bf k}}. 
%\end{align}
%Here $V^0_{\bf q}=\frac{e^2}{2\kappa_0\epsilon_0q}$ is the bare Coulomb
%scattering matrix. 

The scattering term in Eq.~(\ref{ksbe}) can be written as 
\begin{align}\nonumber
\partial_t\rho_{\bf k}(t)|_{\rm{scat}}=& S_{\bf k}(>,<)-S_{\bf k}(<,>)+S_{\bf
  k}^\dagger(>,<)\\ & -S_{\bf k}^\dagger(<,>),
\label{scat-app}
\end{align}
with $S_{\bf k}(>,<)=S_{\bf k}^{\rm ep}(>,<)+S_{\bf k}^{\rm ei}(>,<)+S_{\bf
  k}^{\rm ee}(>,<)$ from the electron-surface optical phonon,
electron-impurity and  electron-electron scatterings, respectively. Here 
\begin{widetext}
\begin{align}
S_{\bf k}^{\rm ep}(>,<)=&\frac{\pi}{\hbar}\sum_{{\bf
    k}^\prime\eta_1\eta_2}|g_{{\bf k}-{\bf k}^\prime}|^2\Lambda_{{\bf
    k},{\bf k}^\prime}\rho_{{\bf k}^\prime}^>(t)T_{\eta_1}\Lambda_{{\bf
    k}^\prime,{\bf k}}T_{\eta_2}\rho_{{\bf k}}^<(t)[N_0^<\delta(\varepsilon_{{\bf k}^\prime\eta_1}-\varepsilon_{{\bf
    k}\eta_2}-\hbar\omega_0)+N_0^>\delta(\varepsilon_{{\bf k}^\prime\eta_1}-\varepsilon_{{\bf
    k}\eta_2}+\hbar\omega_0)],\label{sep}\\
S_{\bf k}^{\rm ei}(>,<)=&\frac{\pi n_i}{\hbar}\sum_{{\bf k}^\prime\eta_1\eta_2}|D_{{\bf
  k}-{\bf k}^\prime}|^2\Lambda_{{\bf k},{\bf k}^\prime}\rho_{{\bf k}^\prime}^>(t)T_{\eta_1}\Lambda_{{\bf
    k}^\prime,{\bf k}}T_{\eta_2}\rho_{{\bf
    k}}^<(t)\delta(\varepsilon_{{\bf k}^\prime\eta_1}-\varepsilon_{{\bf
    k}\eta_2}),\label{sei}\\\nonumber
S_{\bf k}^{\rm ee}(>,<)=&\frac{\pi}{\hbar}\sum_{{\bf k}^\prime\eta_1\eta_2}\Lambda_{{\bf
    k},{\bf k}^\prime}\rho_{{\bf
    k}^\prime}^>(t)T_{\eta_1}\Lambda_{{\bf k}^\prime,{\bf k}}T_{\eta_2}\rho_{{\bf
    k}}^<(t)\sum_{{\bf q}\eta_3-\eta_6}V_{{\bf k}-{\bf
    k}^\prime}^r(\varepsilon_{{\bf q}\eta_3}-\varepsilon_{{\bf q}+{\bf k}^\prime-{\bf
  k}\eta_4})V_{{\bf k}-{\bf k}^\prime}^a(\varepsilon_{{\bf q}\eta_5}-\varepsilon_{{\bf q}+{\bf k}^\prime-{\bf
  k}\eta_6})\\ &\times \mbox{Tr}[\Lambda_{{\bf q},{\bf q}+{\bf k}^\prime-{\bf k}}T_{\eta_6}\rho_{{\bf q}+{\bf k}^\prime-{\bf k}}^<(t)T_{\eta_4}\Lambda_{{\bf q}+{\bf k}^\prime-{\bf k},{\bf
  q}}T_{\eta_3}\rho_{\bf q}^>(t)T_{\eta_5}]\delta(\varepsilon_{{\bf k}^\prime\eta_1}-\varepsilon_{{\bf k}\eta_2}
+\varepsilon_{{\bf q}\eta_5}-\varepsilon_{{\bf q}+{\bf k}^\prime-{\bf k}\eta_6}).\label{see}
\end{align}
\end{widetext}
To obtain the above equations, the Markovian approximation, $\rho^I_{\bf
    k}(t^\prime)\approx\rho^I_{\bf k}(t)$, is adopted in the interaction
  picture. Here $\rho^I_{\bf k}(t)\equiv e^{iE_{\bf k}t}\rho_{\bf
    k}(t)e^{-iE_{\bf k}t}$ with the matrix $E_{\bf
    k}\equiv\mbox{diag}\{\varepsilon_{{\bf  k}+},\varepsilon_{{\bf
      k}-}\}$. Equivalently, in the Schr\"odinger picture, the Markovian
  approximation is written as $\rho_{\bf k}(t^\prime)\approx e^{-iE_{\bf k}(t^\prime-t)}\rho_{\bf
    k}(t)e^{iE_{\bf k}(t^\prime-t)}$. With this approximation, the scattering
  terms of KSBEs\cite{wu-review} are consistent with those of kinetic Bloch
  equations in semiconductors.\cite{haug} However, in the work by Culcer {\sl et
  al.}, the Markovian approximation is mistakenly carried out as $\rho_{\bf k}(t^\prime)\approx \rho_{\bf
    k}(t)$ in the Schr\"odinger picture.\cite{culcer155457} Therefore, the scattering term given there
  deviates from ours. In Eqs.~(\ref{sep})-(\ref{see}), $\Lambda_{{\bf k},{\bf k}^\prime}=U_{\bf k}^\dagger U_{{\bf k}^\prime}$ [$U_{\bf k}$ is given in Eq.~(\ref{uk})] and
$T_{\eta_i}=\frac{1}{2}(1+\eta_i\sigma_z)$ with $\eta_i=\pm$. $\rho_{\bf k}^<(t)\equiv\rho_{\bf k}(t)$ and $\rho_{\bf
  k}^>(t)\equiv1-\rho_{\bf k}(t)$. $N_0^>\equiv N_0+1$ and $N_0^<\equiv N_0$
where $N_0$ is the Boson distribution of surface
optical phonons with energy $\hbar\omega_0=7.4$~meV.\cite{zhu186102} $|g_{\bf
  q}|^2$ is the electron-phonon scattering matrix element given by
Eq.~(\ref{ep-me}). $|D_{\bf q}|^2=|Z_iV_q^0e^{-qd}/\epsilon({\bf q},0)|^2$ is the electron-impurity
scattering matrix element, where $Z_i$ is the charge number of impurity (assumed
to be 1 here), $d$ is the effective distance of impurities from the surface
two-dimensional electron layer (assumed to be zero following Culcer {\sl et
al.}\cite{culcer155457}), and $V_q^0=2\pi\hbar v_f r_s/q$ with 
$r_s=e^2/(4\pi\epsilon_0\kappa_0\hbar v_f)$. In Eq.~(\ref{see}),
$V_{\bf q}^r(\hbar\omega)=V_q^0/\epsilon({\bf q},\omega)$ and $V_{\bf 
  q}^a(\hbar\omega)={V_{\bf q}^r}^\ast(\hbar\omega)$ are the retarded
and advanced Coulomb potentials with dynamic screening, respectively. The RPA
screening $\epsilon({\bf q},\omega)=1-V_q^0\Pi({\bf q},\omega)$,\cite{hwang205418,ramezanali214015,culcer155457} with
\begin{align}\nonumber
  \Pi({\bf q},\omega)=&\sum_{{\bf k}\eta_1\eta_2}\frac{f_{{\bf k}\eta_1}-f_{{\bf
        k}+{\bf q}\eta_2}}{\hbar\omega+\varepsilon_{{\bf
        k}\eta_1}-\varepsilon_{{\bf k}+{\bf
        q}\eta_2}+i0^+}\\ &\times[1+\eta_1\eta_2\cos(\theta_{\bf k}-\theta_{{\bf
        k}+{\bf q}})]/2.
\label{rpa}
\end{align}
In the $n$-type degenerate (intrinsic nondegenerate) situation, under the long-wavelength and static limit,
$\Pi(0,0)=-k_f/(2\pi\hbar v_f)$ (Ref.~\onlinecite{culcer155457}) [$\Pi(0,0)=-k_BT\ln 4/(2\pi\hbar^2
  v_f^2)$] given by Eq.~(\ref{rpa}) with $f_{{\bf 
    k}\eta}$ substituted by the equilibrium
Fermi distribution. Therefore $\epsilon(0,0)=1+q_s/q$ with $q_s=r_sk_f$
($q_s=r_sk_BT\ln 4/\hbar v_f$) for the $n$-type degenerate
(intrinsic nondegenerate) case. Especially, for the intrinsic case, due to the large Fermi
velocity $v_f$ and small $r_s$, the screening is very weak and we further
approximate $\epsilon(0,0)\approx 1$. These approximations for screening are
utilized in analytically calculating the momentum relaxation 
time limited by the electron-impurity scattering in Sec.~\ref{ana1}.

\section{Energy loss rate due to the electron-phonon scattering}
\label{ap2}
The energy loss rate, induced by the electron-phonon scattering solely, satisfies
\begin{align}
\partial_t\varepsilon_o(t)=\sum_{\bf k}\hbar v_fk\mbox{Tr}[\sigma_z\partial_t\rho_{\bf k}(t)\big|_{\rm scat}^{\rm
  ep}].
\end{align}
Near the equilibrium, approximating $\rho_{\bf k}(t)$ isotropically and
retaining its diagonal part only, one has
\begin{align}\nonumber
\partial_t\varepsilon_o(t)&\approx\frac{|g_0|^2}{4\pi\hbar}\sum_{\eta=\pm}\Big\{k_0\int_0^{+\infty}
  dkk(k+k_0)[N_0^>f_{k+k_0}^\eta(t)\\\nonumber
  &-N_0^<f_k^\eta(t)-f_k^\eta(t)f_{k+k_0}^\eta(t)]
  +\int_0^{k_0}dkk^2(k_0-k)\\ & \times[N_0^<(f_{k_0-k}^\eta(t)+f_k^\eta(t)-1)+f_{k_0-k}^\eta(t)f_k^{-\eta}(t)]
\Big\}.
\end{align}
Here $f_{\bf k}^+(t)\equiv f_{{\bf k}+}(t)$ and $f_{\bf k}^-(t)\equiv 1-f_{{\bf
    k}-}(t)$ stand for electron and hole distributions in the conduction and valence bands,
respectively. On the right-hand side of the equation, the first integral is
contributed by the intra-band electron-phonon scattering in both bands while the second one by the
inter-band electron-phonon scattering. For the $n$-type degenerate case with
both low temperature and electron density, we take into
account the contribution of the intra-conduction band electron-phonon
scattering to calculate the energy loss rate as
\begin{align}\nonumber
\partial_t\varepsilon_o(t)\approx&\frac{|g_0|^2k_0^2}{4\pi\hbar}\int_0^{+\infty}
  dkk[N_0^>f_{k+k_0}^+(t)-N_0^<f_k^+(t)\\&-f_k^+(t)f_{k+k_0}^+(t)].
\end{align}  
By using the detailed balance condition satisfied in the initial equilibrium
state, $N_0^>f^+_{k+k_0}(0)-N_0^<f_k^+(0)-f_k^+(0)f_{k+k_0}^+(0)=0$, one has
\begin{align}\nonumber
\partial_t\varepsilon_o(t)\approx&\frac{|g_0|^2k_0^2}{4\pi\hbar}\int_0^{+\infty}
  dkk\Big\{N_0^>[f_{k+k_0}^+(t)-f_{k+k_0}^+(0)]\\\nonumber
  &-N_0^<[f_k^+(t)-f_k^+(0)]-f_k^+(t)f_{k+k_0}^+(t)\\\nonumber &+f_k^+(0)f_{k+k_0}^+(0)\Big\}\\\nonumber
  \approx& \frac{|g_0|^2k_0^2}{4\pi\hbar}\int_0^{+\infty}dkk [f_{k}^+(t)-f_{k}^+(0)]\\=&\frac{|g_0|^2k_0^2}{2\hbar}[n_e(t)-n_e(0)].
\end{align}
This relation is utilized to obtain Eq.~(\ref{nv}).
\end{appendix}

\end{document}